\documentclass[a4paper]{jpconf}
\usepackage{graphicx}
\usepackage{amsmath}
\begin{document}
\title{Flavor SU(3) analysis of charmless $B\to PP$ decays}

\author{Cheng-Wei Chiang}
\address{Department of physics, National Central University, Chungli, Taiwan 320, R.O.C.}
\author{Yu-Feng Zhou}
\address{Theory group, KEK, Tsukuba, 305-0801, Japan}


\begin{abstract}
We perform a global fits to charmless $B \to PP$ decays which
independently constrain the $(\bar\rho,\bar\eta)$ vertex of the
unitarity triangle. The fitted amplitudes and phase are used to
predict the branching ratios and CP asymmetries of all decay modes,
including those of the $B_s$ system.  Different schemes of SU(3)
breaking in decay amplitude sizes are analyzed. The possibility of
having a new physics contribution to $K \pi$ decays is also discussed.
\end{abstract}

Although charmless modes are rare processes, they are very sensitive
to the smallest CKM matrix elements through decay amplitudes and
mixing.
With more modes being observed and measured at higher precisions, it
becomes possible to use purely rare decays to provide a independent
determination of the unitarity triangle vertex $(\bar\rho,\bar\eta)$,
expressed in terms of the Wolfenstein parameters, without reference to
the charmonium modes.  It is therefore interesting to see whether the
charmless $B$ decay data alone also provide a CKM picture consistent
with other constraints and to search for indication of new physics. 
There has been several global fits using flavor isospin, SU(3)
invariant matrix elements \cite{su3fitA} or flavor flow topological 
diagram\cite{Chiang:2003pm}.  
In this talk, we present an updated global $\chi^2$ fits to the available
charmless $B\to PP$ decays using the flavor diagram approach
\cite{Chiang:2006ih}.  The fitting parameters include the
Wolfenstein parameters $A$, $\bar\rho$, and $\bar\eta$, magnitudes of
different flavor amplitudes, and their associated strong phases.  To
take into account SU(3) breaking, we also include breaking factors of
amplitude sizes as our fitting parameters in some fits.  
%
%
%
%
In the present approximation, we consider five dominant types of independent
amplitudes: a ``tree'' contribution $T$; a ``color-suppressed'' contribution
$C$; a ``QCD penguin'' contribution $P$; a ``flavor-singlet'' contribution $S$,
and an ``electroweak (EW) penguin'' contribution $P_{EW}$.  The former four
types are considered as the leading-order amplitudes, while the last one is
higher order in weak interactions.  There are also other types of amplitudes,
such as the ``color-suppressed EW penguin'' diagram $P_{EW}^C$, ``exchange''
diagram $E$, ``annihilation'' diagram $A$, and ``penguin annihilation'' diagram
$PA$.  Due to dynamical suppressions, these amplitudes are ignored in the
analysis.  This agrees with the recent observation of the $B^0 \to K^+ K^-$
decay.



To see the effects of SU(3) symmetry breaking, we consider the following four
fitting schemes in our analysis:
1) exact flavor SU(3) symmetry for all amplitudes;
2) including the factor $f_K/f_\pi$ for $|T|$ only;
3) including the factor $f_K/f_\pi$ for both $|T|$ and $|C|$; and
4) including a universal SU(3) breaking factor $\xi$ for all amplitudes on
  top of Scheme~3.
To reduce the number of parameters, we assume exact flavor SU(3) symmetry for
the strong phases in these fits. In addition to the observables in $B \to PP$ modes, we also include $|V_{ub}| =
(4.26 \pm 0.36) \times 10^{-3}$ and $|V_{cb}| = (41.63 \pm 0.65) \times
10^{-3}$ as our fitting observables.

In the first step, we include only $\pi\pi$, $\pi K$ and $KK$
modes. The four scheme gives $\chi^2/dof=18.9/12,18.0/12,16.4/12$ and
$16.1/11$ respectively. The naive factorization motivated scheme 3 has
the best goodness-of-fit.  The best-fitted values of the parameters in
their 1 $\sigma$ ranges ( amplitudes in units of $10^4$eV) are
\begin{align}
              |T|  &=      0.571^{+0.045}_{-0.040} ,
            & |C|  &=      0.360\pm0.046 ,
         &\delta_C  &=      -49.3\pm9.1 ,
\nonumber \\
              |P|  &=      0.122\pm0.002 ,
        & \delta_P  &=      -17.6\pm2.7 ,
          &  |P_{EW}|  &=      0.011\pm0.001 ,
\nonumber \\
       \delta_{P_{EW}}  &=      -18.7\pm4.0 \ .
\end{align}
\noindent
The results show a nontrivial strong phase $\delta_C$ as well as a large
$C/T\sim 0.63$. The fitted $P_{EW}$ agrees with the SM expectation. However,
the low $S(\pi^0 K_S)$ observed by the experiment is not reproduced, which
is the main source of the inconsistency of the fit.   
In this scheme  the
following results for the weak phases $\alpha$, $\beta$, and $\gamma$
are obtained
\begin{equation}
 \alpha = \left( 83^{+6}_{-7} \right)^\circ,
\beta = \left( 26 \pm 2 \right)^\circ,
\gamma = \left( 72^{+4}_{-5} \right)^\circ \ .
\end{equation}
%
The allowed range for $\bar{\rho},\bar{\eta}$ is given in Fig.1a,
which shows an overall agreement with the other global
fits\cite{CKMfitter}.  However, our results favor a slightly
larger $\gamma$ and the area of the UT.

We further carry out the analysis with the inclusion of modes with
$\eta$ and $\eta'$ in the final state. In all the schemes a large $S$
is found, in particular $S=0.047\pm0.003$ for scheme 3, which is
driven by large $Br(\eta'K)$. Other hadronic amplitudes remain almost
unaffected.  Note that the fit results favor an even larger $\gamma$,
which can be clearly seen from Fig.1b and the following best fitted phase angles
\begin{equation}
 \alpha = \left( 80 \pm 6 \right)^\circ ~,
 \beta = \left( 23 \pm 2 \right)^\circ ~,
 \gamma = \left( 77 \pm 4 \right)^\circ ~ .
\end{equation}
%

Using the fitted parameter, we make predictions for all $B_s$ modes.
In particular, we predict $Br(K^+K^-)=(18.9\pm1.0)\times 10^{-6}$
which on the lower side but still consistent with the latest CDF data
$(24.4\pm 4.8)\times 10^{-6}$. Due to the large $S$ obtained from the fits, we find
large predictions for $Br(\eta'\eta')=(48.3\pm4.1)\times 10^{-6}$ and
$Br(\eta\eta')=(22.4\pm1.5)\times 10^{-6}$.
 
\begin{figure}[h]
\includegraphics[width=0.45\textwidth]{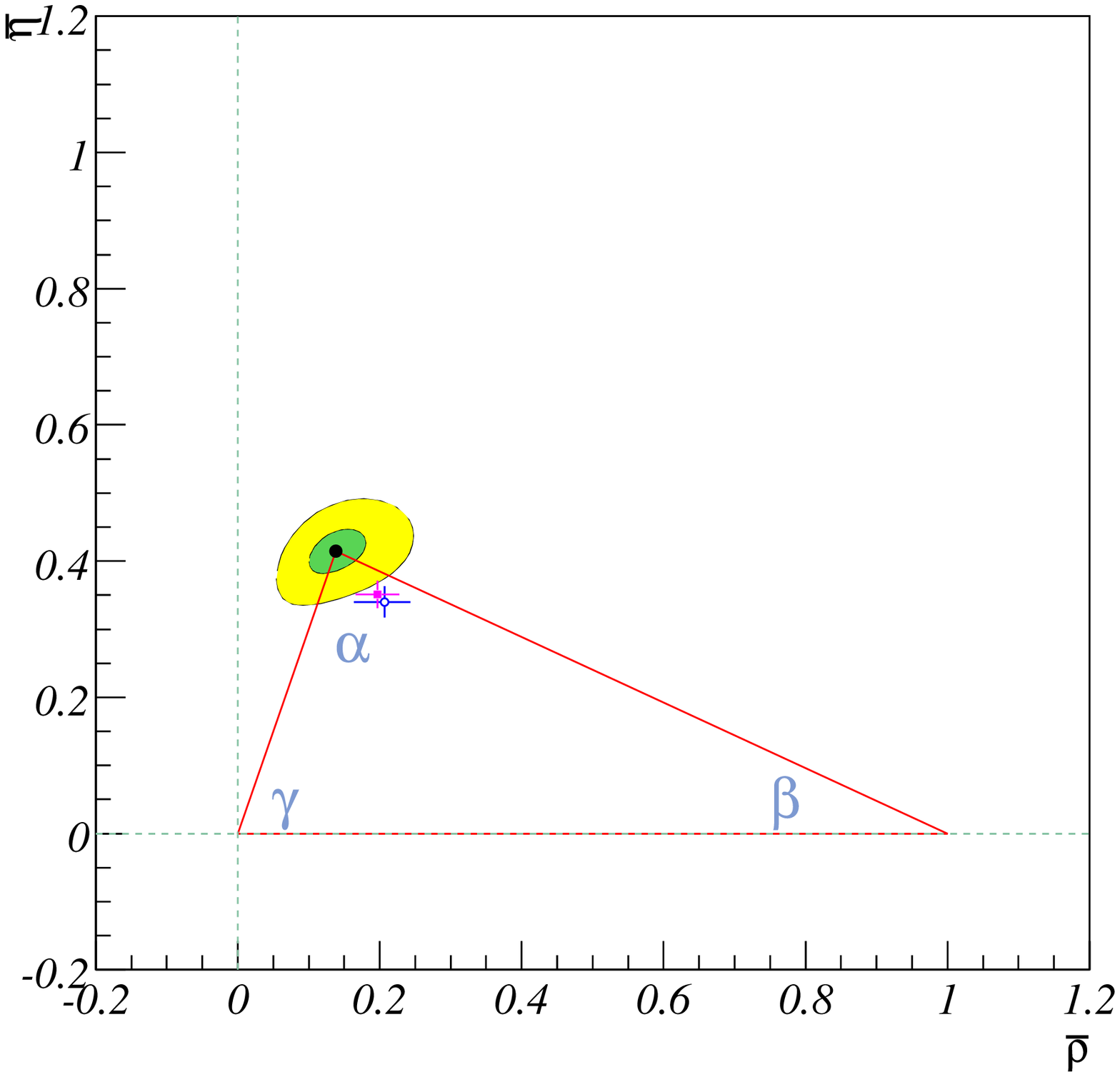}
\includegraphics[width=0.45\textwidth]{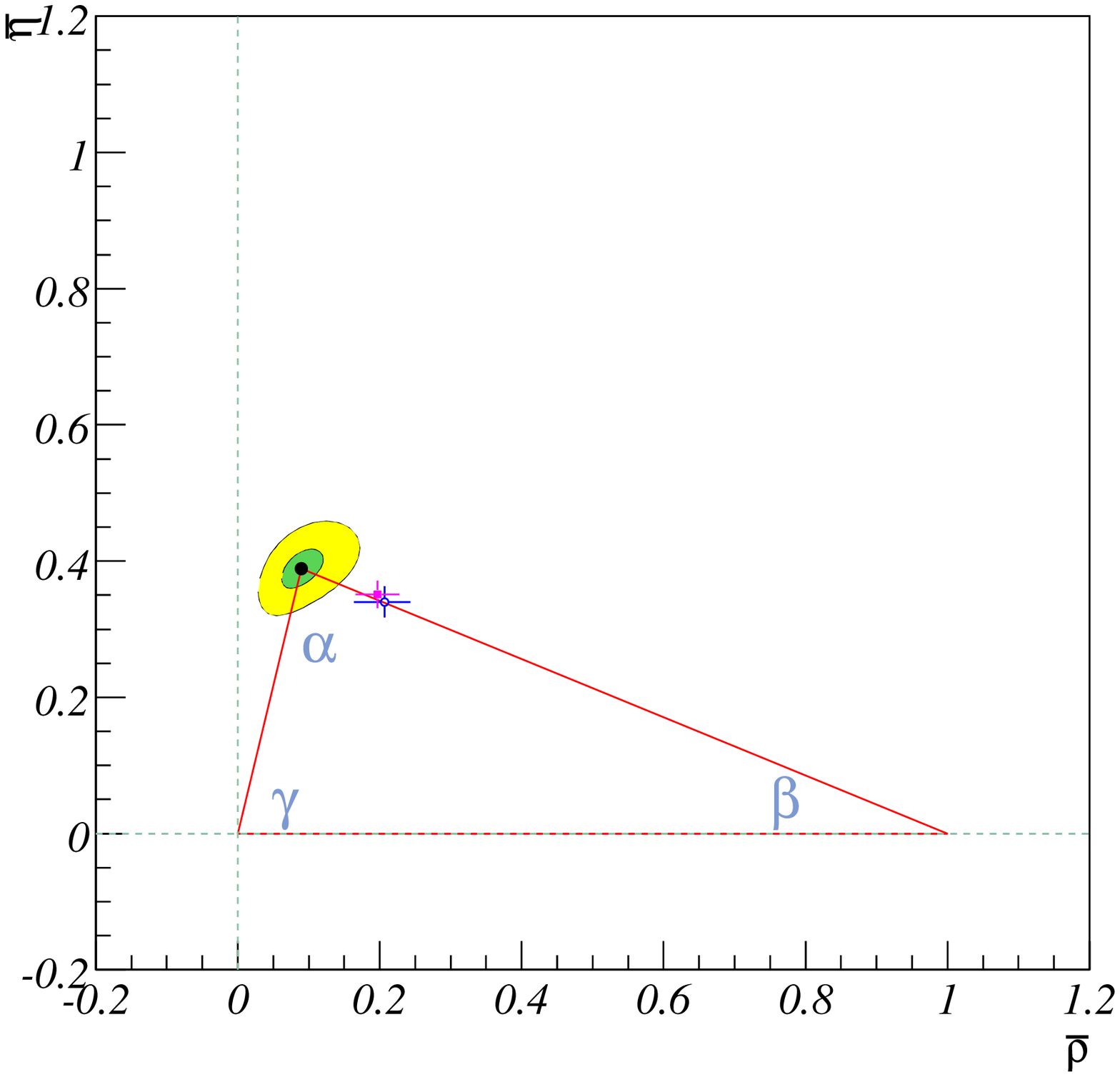}
\caption{
Constraints on the $(\bar\rho,\bar\eta)$ vertex using $B \to \pi \pi,
K \pi$, and $KK$ data in Scheme~3 defined in the text.  Contours
correspond to 1 $\sigma$ and $95\%$ CL, respectively.  The crosses
refer to the 1 $\sigma$ range given by the latest CKMfitter (open
circle) and UTfit (filled square) results using other methods
\cite{CKMfitter} as a comparison. Fig.1a (left) fit with 
$\pi\pi,\pi K$ and $KK$. Fig.1b (right) fit with all $B\to PP$
including $\eta^{(')}$ final states.}  
\label{fig:rhoeta}
\end{figure}

In expectation of possible new physics contributions to the $K \pi$ decays to
account for the observed branching ratio and CP violation pattern
\cite{Yoshikawa,Barger:2004hn,Wu:2004xx,Baek:2006ti}, we try
in Scheme~3 fits with a new amplitude added to these decays.  More explicitly,
a new amplitude $N = |N| \exp{\left[ i (\phi_N + \delta_N) \right]}$ is
included in the $B \to \pi^0 K^-$ and $\pi^0 \bar{K}^0$ decays in such a way
that effectively,
\begin{eqnarray}
  c' \to Y_{sb}^u  C - \left( Y_{sb}^u + Y_{sb}^c \right) P_{EW} + N ~.
\end{eqnarray}
where $Y^q_{q_1 q_2}$ stands for $V^*_{q q_1} V_{q q_2}$.  This
introduces three more parameters ($|N|$, $\phi_N$, and $\delta_N$)
into the fits.  Here we assume that $P_{EW}$ is fixed relative to
$T+C$ through the SM relation.  The $\chi^2_{\rm min}$ is found to
decrease dramatically from $16.4$ to $4.3$ in the limited fit with
only $\pi$, $K$ mesons in the final state.  The new physics parameters
are found to be
\begin{eqnarray}
\label{eq:NPpara}
|N| = 18^{+3}_{-4} \; \mbox{eV} ~, \quad 
\phi_N=(92\pm 4)^\circ ~, \quad \mbox{and} \quad
\delta_N=(-14\pm 5)^\circ ~.
\end{eqnarray}
After rescaled with CKM factors, Such a $|N|$ is about three times as
large as $P_{EW}$ with a large CP violating phase. With this new
amplitudes, the observed low $S(\pi^0 K_S)$ can be accounted for. 
For more detailed discussions on new physics in $P_{EW}$, we refer
to Refs.\cite{Fleischer:2007mq,Imbeault:2006nx}. It
then is of interest to examine if other modes involving $\eta^{(')}$
mesons follow the similar pattern. Our result shows, however, that in
this case the best fitted $N$ is compatible with zero. Possible
reasons for that are $S(\eta'K_s)$ is closer to the SM value, and there
is no $\pi K$ CP puzzle in the $K\eta^{(')}$ modes. In both cases, the
best fitted CKM parameters $\bar\rho$, $\bar\eta$ and $A$ remain
almost unchanged. 

In summary our fits render an area of the $(\bar\rho,\bar\eta)$ vertex
slightly deviated from but still consistent with that obtained from
other constraints. We predict the branching ratios and CP asymmetries
of all decays using flavor SU(3) symmetry, including the $B_s$ system.
The latter will be compared with data already or to be measured at the
Tevatron, large hadron collider (LHC), and KEKB upgraded for running
at $\Upsilon(5S)$.  The possibility of having a new physics
contribution to $K \pi$ decays is  examined from the data fitting
point of view. Although there is some hint of new physics in $\pi K$
system in electroweak penguin sector, it remains to be confirmed from
other $B\to PP$ modes. It would be interesting to see if the current
puzzles in $\pi K$ is correlated to other modes from the future
precision experiments.


\section*{References}

\end{document}